\documentclass[showpacs,twocolumn]{revtex4}

\usepackage{amsmath}
\usepackage{graphicx}

\begin{document}

\title{Inhomogeneous Dust Collapse in 5D Einstein-Gauss-Bonnet Gravity}
\author{S. Jhingan and Sushant G. Ghosh}
\affiliation{Center for Theoretical Physics, Jamia Millia Islamia,
New Delhi 110025, India}

\begin{abstract}
We consider a Lemaitre - Tolman - Bondi type  space-time in Einstein
gravity with the Gauss-Bonnet combination of quadratic curvature
terms, and present exact solution in closed form.  It turns out that
the presence of the  coupling constant of the Gauss-Bonnet terms
$\alpha > 0$ completely changes the causal structure of the
singularities from the analogous general relativistic case. The
gravitational collapse of inhomogeneous dust in the five-dimensional
Gauss-Bonnet extended Einstein equations leads to formation of a
massive, but weak, timelike singularity which is forbidden in
general relativity. Interestingly, this is a counterexample to three
conjecture viz. cosmic censorship conjecture, hoop conjecture and
Seifert's conjecture.
\end{abstract}

\maketitle


\section{Introduction}
The gravitational collapse of an incoherent spherical dust cloud is
described by the metric satisfying Einstein equations $G_{ab} =
\kappa T_{a b}$ with $T_{ab} = \epsilon(t,r) u_{a} u_{b}$, where
$u_a$ is a velocity (i.e. unit-time-like) vector field and
$\epsilon$ is the energy density of the system. This space-time is
described by the Lemaitre-Tolman-Bondi (LTB) solution \cite{tb},
which, in a reference frame $(t,\; r,\; \theta, \; \phi)$ comoving
with the collapsing matter ($u_a = \delta_t^a$), for the marginally
bound case, reads
\begin{eqnarray}\label{tb4d}
ds^2 = & & -dt^2 + \left(\frac{\partial R}{\partial r}\right) dr^2 +
R^2
(d \theta^2 + \sin(\theta)^2 d \phi^2) \nonumber \\
 & & R(t,r) = r \left( 1 - \frac{3}{2}t \sqrt{\frac{2 F(r)}{r^3}}
\right)^{2/3},
\end{eqnarray}
where the mass function $F$, is a function of the coordinate $r$
only, and it is completely determined by the initial state at $t =
0$.  The coordinate $r$ is the comoving radial coordinate, and $t$
is the proper time of freely falling shells. $R$, a function of $t$
and $r$ with $R \geq 0$, is the area coordinate and measures the
actual distance.

This two parameter family solution was a natural extension of the
seminal work of Oppenheimer and Snyder, which led to the
"establishment viewpoint" that the end state of gravitational
collapse is a black hole. It took more than 30 years since the
Oppenheimer-Snyder model \cite{os} till the feasibility of a naked
singularity was brought out by Christodolou \cite{dc} in his study
on the LTB model. It is thus understandable that due to a lack of
any alternative scenario all these years the black hole became the
unique end state of continued gravitational collapse, for a remnant
mass of a collapsing star beyond the threshold neutron star mass
limit. The absence of analytical results led to several conjectures
as well, namely the (weak and strong) cosmic censorship conjecture
(CCC) by Penrose \cite{rp}, hoop conjecture (HC) by Thorne \cite{ks}
and Seifert's \cite{sf} conjecture, which to date remain unproven.

The LTB solution has been extensively used not only to study the
formation of naked singularities and black holes in spherical
collapse, but in cosmology as well. It is well known that the LTB
solutions admit both naked and covered singularities depending upon
the choice of initial data and there is a smooth transition from one
phase to the other \cite{dc,es,rn,tbr,djd,sj,jjs}. These results
have led to strong evidence against the weak CCC \cite{rp}, which
asserts that \emph{there can be no singularity visible from future
null infinity}. In other words, light rays emanate from singularity
but are completely blocked by the event horizon and hence they could
only lay bare singularity to observers who are cofalling with the
collapsing star and not to external observers, while the strong CCC
prohibits its visibility by any observer. That \emph{means no light
rays emanate out of singularity, i.e., singularity is never naked}.
In precise mathematical terms it demands that space-time should be
globally hyperbolic. Despite almost 40 years of effort we are still
far from a general proof of CCC (for recent reviews and references,
see \cite{r1}).

Many studies of gravitational collapse, particularly in cylindrical
symmetry, were also motivated by Thorne's HC of the necessary and
sufficient conditions for the horizon formation: \emph{Horizons form
when and only when a mass M becomes compacted into a region whose
circumference in every direction  $C \leq 4 \pi M$} \cite{ks}. Thus,
planar or cylindrical matter will not form a black hole (black plane
or black string) \cite{ks}. Unlike CCC, the HC does not suffer from
counterexamples. The HC was originally given for four-dimensional
(4D) space-times in general relativity. It was modified for higher
dimensions by Ida and Nakao  \cite{yni}: \emph{Black holes with
horizons form when and only when a mass M gets compacted into a
region whose ($D-3$)dimensional area $V_{D-3}$ in every direction is
$V_{D-3} \leq G_D M$,} where $G_D$ is the gravitational constant in
the D-dimensional theory of gravity, and the ($D-3$)- dimensional
area means the volume of the($D-3$)-dimensional closed submanifold
of a spacelike hypersurface. The HC is related to the trapped
surface conjecture of Seifert \cite{sf} that \emph{massive
singularities have to be trapped.} It should be interesting to see
if these conjectures still hold in Einstein gravity with the
Gauss-Bonnet combination of quadratic curvature terms or in a higher
dimensional space-time.

Current experimental results involving tests of the inverse square
law do not rule out extra dimensions even as large as a tenth of a
millimeter. It is important to consider the evolution of the extra
dimensions since the observed strength of the gravitational force is
directly dependent on the size of the extra dimensions. As a
consequence, there is a renewed interest towards an understanding of
the general relativity in more than four dimensions, as a growing
volume of recent literature indicates.  In particular, several
solutions to the Einstein equations of localized sources in higher
dimensions have been obtained in the recent years \cite{ft,rcm},
from viewpoint of gravitational collapse \cite{hd} and in particular
LTB-like solutions \cite{gab,bsc,gb,gds,gds1}.

In recent years a renewed interest has grown in higher order
gravity, which involves higher derivative curvature terms.   Among
the higher curvature gravities, the most extensively studied theory
is the so-called Einstein-Gauss-Bonnet (EGB) gravity. The EGB
gravity is a special case of Lovelock�s theory of gravitation, whose
Lagrangian contains just the first three terms.  The Gauss-Bonnet
term yields nontrivial dynamics in dimensions greater than or equal
to $5$. It appears naturally in the low-energy effective action of
heterotic sting theory \cite{Gross}. Boulware and Deser \cite{bd}
found exact black hole solutions in $N (\geq 5)$-dimensional
gravitational theories with a four dimensional Gauss-Bonnet term
modifying the usual Einstein-Hilbert action. These solutions are
generalizations of the N-dimensional spherically symmetric black
hole solution found by Tangherlini \cite{ft}, and Myers and Perry
\cite{rcmp}. Other spherically symmetric black hole solutions in the
Gauss-Bonnet gravity have been found and discussed in
\cite{jtw,tmgb,rcm1}, and topologically nontrivial black holes have
been studied in \cite{rgc}.  The effects of Gauss-Bonnet terms on
the Vaidya solutions have been investigated in \cite{tk,hm,gd,dg},
and on the LTB solutions in \cite{maeda}. These papers show that the
appearance of a Gauss-Bonnet term in the field equations has no
effect on the occurrence of locally naked singularity, while it has
some effects on the strength of the curvature.

Recently, Maeda \cite{maeda} considered the spherically symmetric
gravitational collapse of a inhomogeneous dust with the $N (\geq
5)$-dimensional action including the Gauss-Bonnet term. He
investigated its effects on the final fate of gravitational
collapse without finding the explicit form of the solution. In
this paper, we consider the 5D  action with the Gauss-Bonnet terms
for gravity and give a {\em exact model} of the gravitational
collapse of a inhomogeneous dust including the second order
perturbative effects of quantum gravity. A 5D space-time is
particularly relevant because both 10D and 11D supergravity
theories yield solutions where a 5D space-time results after
dimensional reduction \cite{js}.

This paper is organized as follows. In the next section, we derive
the general solutions, in a closed form, for marginally bound case,
which is a kind of generalized LTB space-time in the 5D EGB gravity
with the energy-momentum tensor of a dust. For definiteness we shall
call it 5D-LTB-EGB. The nature of singularities of such a space-time
in terms of its being hidden within a black hole, or whether it
would be visible to outside observers, and the consequence of EGB on
5D-LTB collapse are analyzed in Sec. III. The detailed analysis on
apparent horizon is a subject of Sec. IV, Sec. V is devoted to
strength of singularity, and  is followed by a discussion.

We have used units which fix the speed of light and the
gravitational constant via $8\pi G = c^4 = 1$.

\section{5D Lemaitre-Tolman-Bondi Solutions in Einstein Gauss-Bonnet Gravity}
We begin with the following $5D$ action:
\begin{equation}\label{action}
S = \int d^5 x \sqrt{-g} \left[\frac{1}{2\kappa_5^2}(R + \alpha
L_{GB})\right] + S_{matter},
\end{equation}
where $R$ is a $5D$ Ricci scalar and $\kappa_5 \equiv \sqrt{8\pi
G_5} $ is $5D$ gravitational constant. The Gauss-Bonnet Lagrangian
is of the form
\begin{equation}\label{EGB}
L_{GB} = R^2 - 4 R_{ab}R^{ab}+R_{abcd} R^{abcd},
\end{equation}
where $\alpha$ is the coupling constant of the Gauss-Bonnet terms.
This type of action is derived in the low-energy limit of heterotic
superstring theory~\cite{Gross}. In that case, $\alpha$ is regarded
as the inverse string tension and positive definite and we consider
only the case with $\alpha \ge 0$ in this paper. In the 4D
space-time, the Gauss-Bonnet terms do not contribute to the field
equations. The action (\ref{action}) leads to the following set of
field equations:
\begin{equation}\label{FE}
{\cal G}_{a b} \stackrel{\rm def}
     = G_{ab} + \alpha H_{ab} = T_{a b},
\end{equation}
where
\begin{eqnarray}\label{equations}
  G_{ab} &=& R_{ab} -\frac{1}{2} g_{ab} R
\end{eqnarray}
is the Einstein tensor and
\begin{eqnarray}
  H_{ab} &=& 2[RR_{ab}-2R_{a\alpha}R^{\alpha}_b -
  2 R^{\alpha \beta}R_{a\alpha b\beta} + R_a^{\alpha\beta\gamma}
  R_{b\alpha\beta\gamma}]  \nonumber \\& & -\frac{1}{2}g_{ab}L_{GB}
\end{eqnarray}
is the Lanczos tensor.

The standard LTB solution (\ref{tb4d})
 represents an interior of a collapsing inhomogeneous dust
sphere. The solution we seek is  collapse of a spherical dust in
5D-EGB. The energy-momentum tensor for dust is
\begin{equation}
T_{ab} =  \epsilon(t,r) \delta_{a}^t \delta_{b}^t, \label{eq:emt}
\end{equation}
where $u_a = \delta_t^a$ is the $5D$ velocity. The metric for the
$5D$ case, in comoving coordinates, is  \cite{gab,bsc,gb,gds}:
\begin{equation}\label{metric}
ds^2 = -dt^2 + A(t,r)^2 dr^2 + R(t,r)^2 d\Omega_3^2,
\end{equation}
where $d\Omega_3^2 = (d\theta^2 +\sin^2\theta( d\phi^2 + \sin^2\phi
d\psi^2))$, is a metric on three-sphere. The coordinate $r$ is the
comoving radial coordinate, $t$ is the proper time of freely falling
shells, $R$ is a function of $t$ and $r$ with $R \geq 0$, and $A$ is
also a function of $t$ and $r$. For the metric (\ref{metric}), with
energy-momentum tensor (\ref{eq:emt}), the Einstein field equations
take the form
\begin{eqnarray}\label{fe1}
{\cal G}^{t}_t & = &  \frac{12({R'}^2-A^2(1+{\dot R}^2))}{R^3
A^5}[{R'} {A'}+A^2 {\dot R} {\dot A} - A {R''}] \alpha \nonumber \\
&&- \frac{3}{A^3 R^2} [A^3(1+{\dot R}^2) + A^2 R {\dot R} {\dot A} +
R {R'} {A'} \nonumber \\  &&- A (R {R''}+{R'}^2)] = -\epsilon(t,r),
\\\label{fe2} {\cal G}^r_r & = & -12\alpha\left(\frac{1}{R^3} -
\frac{{R'}^2}{A^2R^3} + \frac{{\dot R}^2}{R^3} \right){\ddot R} +
3\frac{{R'}^2}{A^2R^2} \nonumber \\  & & - 3 \frac{1 + {\dot R}^2
+ R {\ddot R}}{R^2}  = 0, \\ \label{fe3} {\cal
G}^{\theta}_{\theta}& = & {\cal G}^{\phi}_{\phi}={\cal
G}^{\psi}_{\psi} = \frac{4 \alpha}{A^4R^2}\left[
-2A({A'}{R'}+A^2{\dot A}{\dot R}-A{R''}){\ddot R}  \right.
\nonumber
\\ && \left. + A ({R'}^2-A^2(1+ {\dot R}^2)) {\ddot A} +2({\dot
A}{R'}-A{\dot R'})
\right ] \nonumber \\
& & -\frac{1}{A^3R^2}[A^3(1+{\dot R}^2+2R {\ddot R})+A^2R(2{\dot
R}{\dot A}+R{\ddot A}) \nonumber \\ & & +
2R{R'}{A'}-2A(R{R''}+{R'}^2)] = 0,
\\ \label{fe4} {\cal G}^t_r & =& \frac{12\alpha}{A^5 R^3}({\dot A}{R'}-A
{\dot R'})(A^2(1+{\dot R}^2)-{R'}^2) \nonumber \\ &&-3\frac{A{\dot
R'}-{\dot A}{R'}}{A^3 R} =0,
\end{eqnarray}
where an over-dot and prime denote the partial derivative with
respect to $t$ and $r$, respectively. Since space-time is
nonradiating Eq.~(\ref{fe4}), leads to two families of solutions
\begin{equation}
A(t,r) = \frac{R'}{W}, \label{eq:el}
\end{equation}
and
\begin{equation}\label{case2}
A(t,r) = \pm \frac{2 \sqrt{{\alpha}}{R'}}{[{ R^{2}+4{\alpha}({\dot
R}^{2} +1)}]^{1/2}} ,
\end{equation}
where $W= W(r)$ is an arbitrary function of $r$. Note the striking
similarity of function $A$ with analogous 5D-LTB solutions
\cite{gab,bsc,gds}. In what follows we shall consider the case
$A(t,r) = {R'}/{W}$, since in the other case the $\alpha
\rightarrow 0$ leads to a trivial solution.

It is straightforward to check that if the ${\cal G}^r_r=0$
condition is satisfied the other Einstein equations  (namely, ${\cal
G}^{\theta}_{\theta} ={\cal G}^{\phi}_{\phi} = {\cal
G}^{\psi}_{\psi} = 0 $) are automatically satisfied.  Finally, in
order for ${\cal G}^r_r=0$ to hold it is necessary that $R$
satisfies
\begin{equation}
\frac{\ddot{R}}{R} = \frac{\dot{R}^2 - (W^2-1)}{4 \alpha
(W^2-1-\dot{R}^2 ) - R^2},
\end{equation}
which can be easily integrated to yield
\begin{equation}
\dot{R}^2 \left[1 - 4 \alpha\frac{W^2-1}{R^2} \right] = (W^2 - 1) +
\frac{F}{R^2} - 2 \alpha \frac{\dot{R}^4}{R^2}.\label{eq:fe}
\end{equation}
This is the master equation of the system. Here $F = F(r)$ is an
arbitrary function of $r$ and is referred to as mass function.
Substituting Eqs.~(\ref{eq:el}) and (\ref{eq:fe}) into
Eq.~(\ref{fe1}) we obtain
\begin{equation}
F' = \frac{2}{3} \epsilon R^3 R'. \label{eq:m}
\end{equation}
Integrating Eq.~(\ref{eq:m}) leads to
\begin{equation}
F(r) = \frac{2}{3} \int \epsilon R^3 dR, \label{eq:m1}
\end{equation}
where the constant of integration is taken as zero since we want a
finite distribution of matter at the origin $r = 0$. We note that
$F'$ (as well as $F$) must be positive.  Indeed the energy density,
$\epsilon$, must be non-negative.  It is easy to see that as $\alpha
\rightarrow 0$ the master solution (\ref{eq:fe}) of the system
reduces to the corresponding 5D-LTB solution in \cite{gab,gds}
\begin{equation}
\dot{R}^2 = W^2 - 1 + \frac{{F}}{R^2}. \label{eq:fe1}
\end{equation}

It may be noted that in a general relativistic case ($\alpha
\rightarrow 0$), Eq.~(\ref{eq:fe}) [or  (\ref{eq:fe1})] has three
types of solutions, namely, hyperbolic, parabolic and elliptic
solutions depending on whether $W(r) > 1$, $W(r) = 1$ or $W(r) < 1$
respectively \cite{gab,gds,gds1}.  Analogously, here the condition
$W(r)=1$, is the marginally bound condition, meaning collapsing
shell is at rest at spatial infinity $(R=\infty)$. From
Eq.~(\ref{eq:fe}), we obtain
\begin{eqnarray}\label{dyn0}
{\dot R}^2 &=& (W^2-1) -  \frac{R^2}{4 \alpha} \nonumber \\
&\mp& \frac{R^2}{4 \alpha} \left[{1+ \frac{16
\alpha^2}{R^4}(W^2-1)^2 + \frac{8 \alpha F(r)}{R^4}} \right]^{1/2}.
\end{eqnarray}

There are two families of solutions which correspond to the sign in
front of the square root in Eq.~(\ref{dyn0}). We call the family
which has the minus (plus) sign the minus  (plus) branch solution.
In the general relativistic limit ${ \alpha} \to 0$,  we recover the
$5D$-LTB solution in Einstein gravity \cite{gb,gds}. There is no
such limit for the plus-branch solution.  We consider the
minus-branch solution in order to compare with general relativistic
case. Eq.~(\ref{dyn0}) is a modified {Friedmann}-like equation in
5D-EGB and is a bit complicated compared to the corresponding
general relativistic Eq.~(\ref{eq:fe1}).  Maeda \cite{maeda} has
analyzed LTB models near the center ($r \sim 0$) in EGB without
finding an explicit solution.  Here we present the 5D-LTB-EGB exact
solution in closed form, which facilitates us to
  explicitly analyze the final fate of gravitational collapse.

Henceforth, we shall confine ourselves to the marginally bound case
$[W(r)=1]$. In the present discussion, we are concerned with
gravitational collapse, which requires $\dot{R}(t,r) < 0$. Eq.
(\ref{dyn0}) can be integrated to
\begin{eqnarray}\label{solution}
{t_\varsigma(r)-t}  = {\frac{\sqrt{\alpha}}{2\sqrt{2}}} \tan^{-1}
\left[\frac{3 R^2 -\sqrt{R^4+8\alpha F}}{2\sqrt{2} R [\sqrt{R^4 +
8\alpha F} -R^2]^{1/2}} \right] \nonumber \\ +\sqrt{\frac{\alpha
R^2}{\sqrt{R^4+8\alpha F}-R^2}}\;, \quad
\end{eqnarray}
where $t_\varsigma(r)(r)$ is an arbitrary function of integration.
 As it is possible to make an arbitrary
relabeling of spherical dust shells by $r \rightarrow g(r)$, without
loss of generality, we fix the labeling by requiring that, on the
hypersurface $t = 0$, $r$ coincide with the area radius
\begin{equation}
R(0,r) = r.             \label{eq:ic}
\end{equation}
This corresponds to the following choice of $t_\varsigma(r)$:
\begin{eqnarray}\label{scale}
{t_\varsigma(r)} = {\frac{\sqrt{\alpha}}{2\sqrt{2}}} \tan^{-1}
\left[\frac{3 - \sqrt{1 + 8\alpha {\tilde F}}}{2\sqrt{2}[\sqrt{1 +
8\alpha {\tilde F}}-1]^{1/2}}\right] \nonumber \\ + \sqrt{
\frac{\alpha}{\sqrt{1+8\alpha {\tilde F}}-1}}\; , \quad
\end{eqnarray}
where $\tilde F = F/r^4$.

In order to study the collapse of a finite spherical body in EGB, we
have to match the solution along the time-like surface at some $r =
r_c >0 $ to the $5D$-EGB Schwarzschild exterior discovered by
Boulware and Desser \cite{bd} and Wheeler \cite{jtw}. On a spherical
hypersurface $\Sigma$, the junction conditions yield $F = M_s$
\cite{maeda}.  Here $r_s = R(t,r_c)$, and $M_s$ is the total mass
enclosed within the coordinate radius $r_c$ \cite{maeda}.

Also, in the general relativistic case the  energy-momentum tensor
given by Eq.~(\ref{eq:emt}) satisfies the weak energy condition.
It means that the energy density as measured by any local observer
is non-negative.  However, this may  true in EGB because the
Gauss-Bonnet term itself violates the energy condition (like a
negative cosmological constant).

\section{Initial Data and singularity}
The final fate of gravitational collapse and the nature of the
 singularity continues to be among one of the most outstanding
problems in general relativity. As pointed out earlier, the
conjecture that such a collapse, leading to a singularity, under
physically realistic conditions must end in the formation of a black
hole, and that the eventual singularity must be covered by an event
horizon is the CCC. Despite numerous attempts, this conjecture as
such remains a major unsolved problem lying at the foundation of
black hole physics today. From such a perspective, it is worthwhile
to examine the nature of the singularity, in terms of its visibility
for an observer, when it develops in the context of the 5D-LTB-EGB.
In LTB space-time, shell crossing singularities are defined by
$R'=0$ and they can be naked. It has been shown in LTB case
\cite{rn} that shell crossing singularities are gravitationally weak
and hence such singularities cannot be considered seriously in the
context of the CCC.  On the other hand, in general relativity
central shell focusing singularities (characterized by $R=0$) can
also be naked and gravitationally strong as well. Thus, unlike shell
crossing singularities, shell focusing singularities do not admit
any metric extension through them and are considered to be the
genuine singularities of space-time. This led us  to investigate a
similar situation in 5D-LTB-EGB space-time, because shell crossing
singularities  are assumed to be extendible in general relativity.
Christodoulou \cite{dc} pointed out in the LTB case that the
noncentral singularities cannot be naked. This is also true for all
spherically symmetric models (including models in higher dimensions)
for physically reasonable matter fields. It will be interesting to
discuss if this feature gets modified by introduction of
 the Gauss-Bonnet term.

The easiest way to detect a singularity in a space-time is to
observe the divergence of some invariants of the Riemann tensor. The
Kretschmann scalar ($\mbox{K} = R_{abcd} R^{abcd}$, $R_{abcd}$ is
the Riemann tensor) for the metric (\ref{metric}) with the help of
(\ref{eq:el}) reduces to
\begin{equation}
K = 12 \frac{\ddot{R}^2}{R^2} + 12 \frac{\dot{R}^4}{R^4} + 4
\frac{\ddot{R}'^2}{R'^2} + 12 \frac{\dot{R}^2\dot{R}'^2}{R'^2R^2}
\label{eq:ks}
\end{equation}
It can be verified that the Kretschmann scalar is finite on the
initial data surface. For our case the general expression for energy
density is
\begin{equation}\label{e-dens}
\epsilon(t,r) = \frac{3F'}{2R^3 R'}.
\end{equation}
Hence, it is clear that if $F'$ is regular and bounded away from
zero, then energy density diverges when   ${R'} = 0$ and $R = 0 $.
Hence, we have both shell crossing as well as shell focusing
singularities for, respectively, ${R'} = 0$ and $R = 0 $. For $t =
t_c(r)$, we have $R(t,r) = 0$, which is the time when the matter
shell $r$ = constant hits the physical singularity.  Further, the
Kretschmann scalar diverges at $t=t_{c}(r)$ indicating the presence
of a scalar polynomial curvature singularity \cite{he}.

The (shell focusing) singularity curve can be obtained using Eq.
(\ref{solution}) as
\begin{equation}\label{sing-curve}
t_c(r) =t_\varsigma(r) +\frac{\pi \sqrt{\alpha}}{4\sqrt{2}},
\end{equation}
which represents the proper time for the complete collapse of a
shell with coordinate $r$. Interestingly, positive $\alpha$ delays
the formation of singularity. In the limit of vanishing $\alpha$ we
recover the crunch time for relativistic 5D-LTB. The two arbitrary
functions $F(r)$ and $t_{c}(r)$ completely specify the dynamics of
collapsing shells.

Analogous to LTB models, in the case of positive $\alpha$, the
evolution always leads to formation of a shell focusing curvature
singularity. The mass function $F$ can be related with initial data
(density) at the scaling surface, $t=0$ $(R=r)$, where
(\ref{e-dens}) reduces to form
\begin{equation}\label{initial}
F(r) = \frac{2}{3}\int_0^{r} \epsilon(0,r) r^3 dr ,
\end{equation}
which completely specifies the mass function in terms of the initial
density profile. The function $F$ must be positive, because $F < 0 $
implies the existence of negative mass. This can be seen from the
mass function $m(t,r)$ \cite{missharp, gab},
 which in the 5D-LTB-EGB case is given by
\begin{eqnarray}
m(t,r) & = & R^2 \left(1 - g^{ab} R_{,a} R_{,b} \right) \nonumber\\
& = &R^2 \left(1 - \frac{{R'}^2}{A^2} + \dot{R}^2 \right).
\label{eq:m2}
\end{eqnarray}
Using Eqs.~(\ref{eq:el}) and ~(\ref{eq:fe}) into Eq.~(\ref{eq:m2})
we get
\begin{equation}
m(t,r) = F(r)  - 2 \alpha \dot{R}^4.\label{eq:m3}
\end{equation}
It may noted that one can also calculate mass using the formula
proposed by Maeda \cite{maeda,QLMaeda} for the generalized mass
function in the EGB. The mass function $F(r) = m(t,r) + 2 \alpha
\dot{R}^4$, is equivalent, up to a constant factor, to the
generalized mass function in EGB \cite{maeda,QLMaeda}.

Next, we study the structure of singularities in 5D-LTB-EGB
space-time and compare it with the general relativistic case by
using solution obtained in the previous section. Consider a
spherically symmetric dust cloud with density profile:
\begin{equation}\label{initdata}
\epsilon (0,r) = \epsilon_0 \left[1 - \left(\frac{r}{r_b}\right)^n
\right] .
\end{equation}
Here, $\epsilon_0$ is the central density and $r_b$ is the boundary
of the collapsing cloud. This is a profile where energy density
decreases as we move away from the center, as is expected inside a
star. Initial data for a model are completely specified by
$\epsilon_0, \; r_b, \; n$ and $\alpha$. Since matter is
pressureless fluid matching to a suitable exterior requires matching
of the mass function of the interior and exterior space-times. We
would like to mention here that, like LTB models \cite{sj,jjs}, we
can choose an arbitrary profile of the form $\epsilon(0,r) =
\epsilon_0 + \epsilon_1 r + \cdots$. However, as it is well known
\cite{sj,jjs} only the first nonvanishing term in the density
gradient is important in deciding the causal structure of a
singularity near the center. Hence our choice of density profile in
no way restricts the generality of our analysis.

The mass function corresponding to the density profile given above
is of the form:
\begin{equation}\label{mass-func}
F(r) = \frac{\epsilon_0}{6} r^4 \left[1-\frac{4 r^n}{(n+4)r_b^n}
\right] .
\end{equation}
One of the important ingredients in singularity theorems is the
assumption of trapped surfaces. The important issue in collapse is
to show whether such trapped surfaces form during collapse or not.
More importantly there should not be {\em a priori} trapped surfaces
present in the initial data surface. The condition for the existence
of an apparent horizon (the inner boundary of the region containing
trapped surfaces), two-spheres with outward normals as null, is
\begin{equation}\label{app}
g^{\mu \nu} R_{,\mu} R_{,\mu} = - \dot{R}^2 + \frac{R{'}^2}{A^2}= 0
\end{equation}
Demanding the absence of trapped surfaces (\ref{app}) in initial
data implies,
\[
g^{\mu \nu} R_{,\mu} R_{,\mu}|_{t=0} > 0.
\]
For the mass function of the form (\ref{eq:m2}) this condition
reduces to
\[
r^2 \left[ -1 +
\sqrt{1+\frac{4\alpha\epsilon_0}{3}\left(1-\frac{4r^n}{(n+4)r_b^n}\right)}\right]
< 4\alpha.
\]
Therefore, for a given density profile and central density the size
of the cloud is limited by the magnitude of parameter $\alpha$.
Note, however, that collapsing matter being dust the density need
not vanish at the boundary for a smooth matching to the vacuum
exterior.

\section{Dynamics of Apparent Horizon}
One of the important constructions in general relativity is that of
a trapped surface. In a 4D space-time, it is a compact $2D$, smooth
spacelike submanifold with the property that the expansion of future
directed null geodesics (outgoing as well as ingoing), orthogonal to
this submanifold, is negative everywhere ~\cite{rmw}. They are
crucial in proving null-geodesic incompleteness in context of
gravitational collapse. The apparent horizon (AH) is the outermost
marginally trapped surface for the outgoing photons.  The AH can be
either null or spacelike, that is, it can 'move' causally or
acausally \cite{hayward}.  The main advantage of working with the
apparent horizon is that it is local in time and can be located at a
given spacelike hypersurface. Moreover, even if energy conditions
hold the whole scenario of the event horizon still remains unclear
in EGB \cite{maeda2}.

Considering Eq. (\ref{eq:fe}), the apparent horizon condition
(\ref{app}) becomes
\begin{equation}\label{app-cond}
R(t_{AH}(r),r) =  \sqrt{F(r) - 2 \alpha}.
\end{equation}
It is clear that the presence of  the coupling constant of the
Gauss-Bonnet terms $\alpha$ produces a change in the location of
these horizons. Such a change could have a significant effect in the
dynamical evolution of these  horizons. In the relativistic limit,
$\alpha \rightarrow 0$,  $R_{AH} \rightarrow \sqrt{F(r)}$ \cite{gb}.
For nonzero $\alpha$ the structure of the apparent horizon is
non-trivial. Interestingly the theory demands $\alpha$ to be a
positive number which forbids apparent horizon from reaching the
center thereby making the singularity massive and eternally visible,
which is forbidden in the corresponding general relativistic
scenario. In general relativity noncentral singularity is always
covered \cite{dc} (see also \cite{cjjs}). However, in the presence
of the Gauss-Bonnet term we find that even the noncentral
singularity is naked, in spite of being massive [$F(r>0]
> 0$). Further, Eq.~(\ref{app-cond}) has a mathematical
similarity for the analogous situation in null fluid collapse where
the expression for the apparent horizon is $r_{AH} =  \sqrt{m(v) - 2
\alpha}$ \cite{gd}.

\begin{figure}[ht]
\centering
\includegraphics[width=6.0 cm,angle=-90]{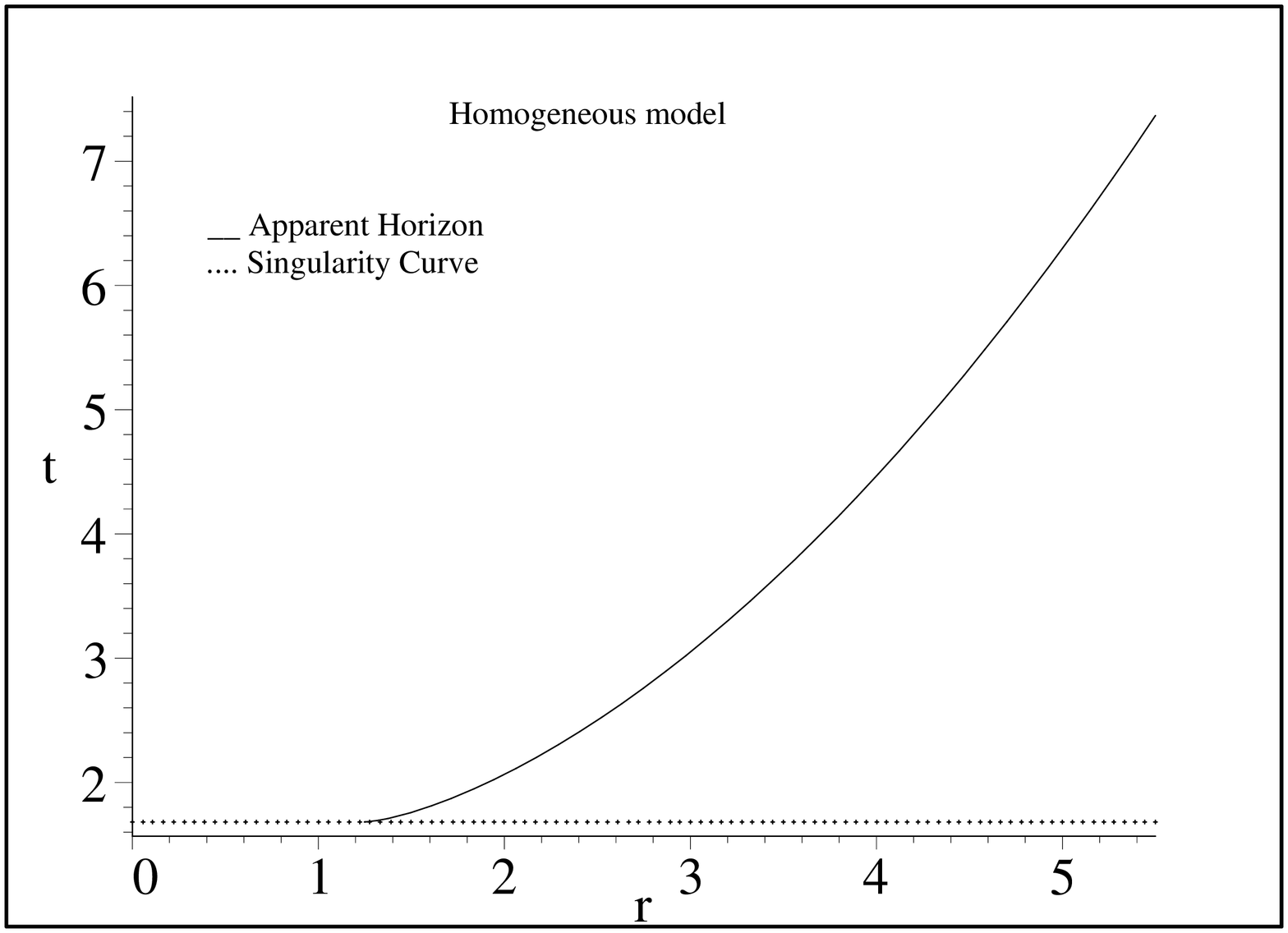}
\includegraphics[width=6.0 cm,angle=-90]{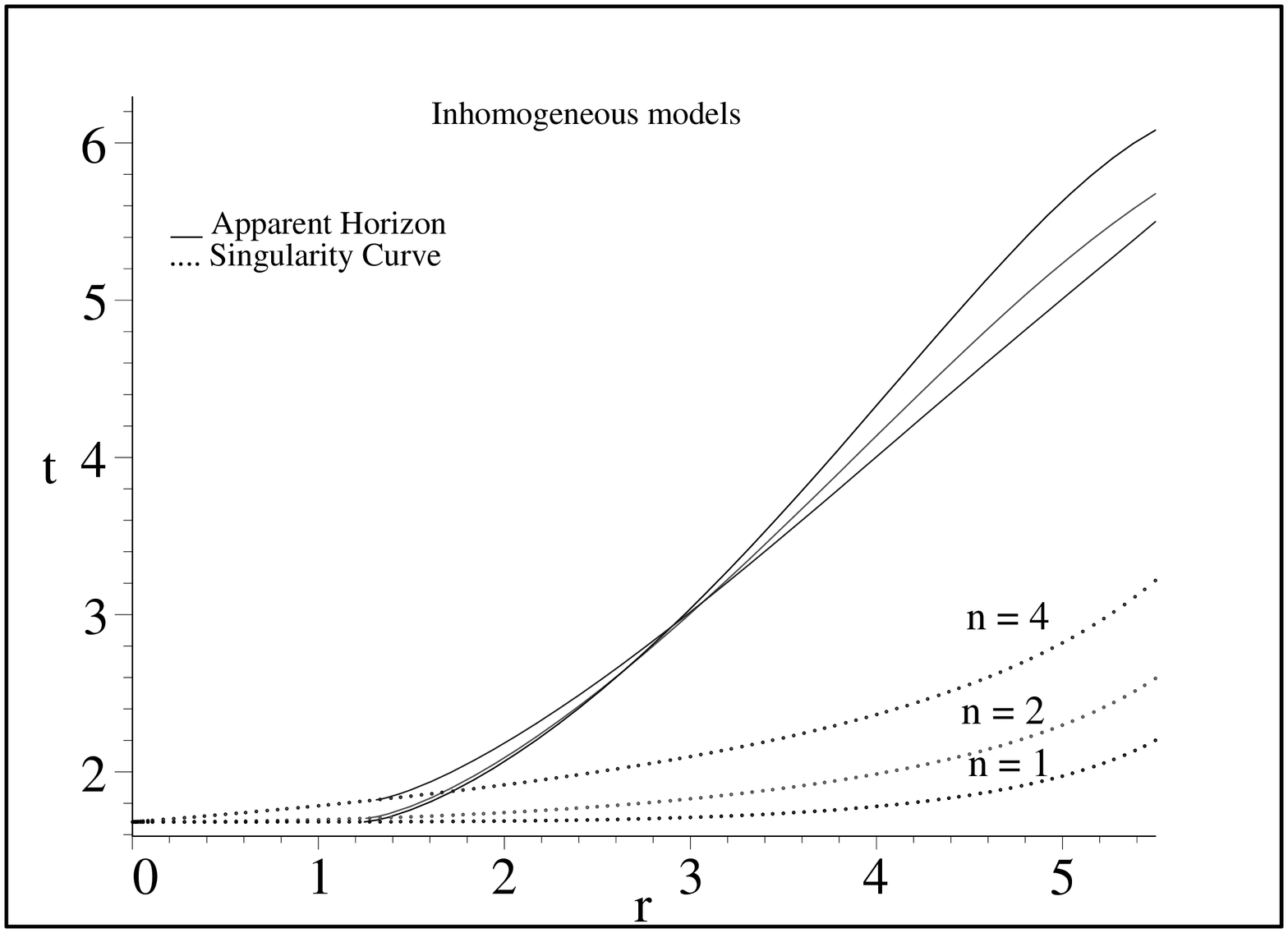}
\caption{A 2D picture of the EGB collapse showing formation of
singularity and apparent horizon for the parameter values
$\epsilon_0 = 1, \alpha = 0.2, r_b = 5.5$. The dotted curves
represent singularity whereas the apparent horizons are the
continuous one }\label{figure1}
\end{figure}

Equation~(\ref{app-cond}) implicitly defines a curve $t_{ah}(r)$
and represents the apparent horizon, i.e. the time at which the
shell gets trapped. Since the collapse is spherical the whole
framework can be expressed by a 2D picture, where the singularity
curve Eq. (\ref{sing-curve}) represents the time of complete
collapse of the shell labeled $r$. To further analyze the horizon
curve, we combine Eqs. (\ref{solution}) and (\ref{sing-curve})
giving
\begin{eqnarray}\label{new_dynamics}
{t_c(r)-t} &=& \frac{\pi\sqrt{\alpha}}{4\sqrt{2}} +
\sqrt{\frac{\alpha R^2}{\sqrt{R^4+8\alpha F}-R^2}} \\
\nonumber &+&{\frac{\sqrt{\alpha}}{2\sqrt{2}}} \tan^{-1}
\left[\frac{3 R^2 -\sqrt{R^4+8\alpha F}}{2\sqrt{2} R [\sqrt{R^4 +
8\alpha F} -R^2]^{1/2}} \right] .
\end{eqnarray}
Then, the apparent horizon condition (\ref{app-cond}) reduces
Eq.~(\ref{new_dynamics}) to form
\begin{eqnarray}\label{sin-app}
t_c(r) - t_{AH}(r) = \frac{\pi \sqrt{\alpha}}{4\sqrt{2}}
+\frac{\sqrt{\alpha}}{2\sqrt{2}} \tan^{-1}\left[
\frac{F-4\alpha}{2\sqrt{2\alpha (F-2\alpha)}} \right] \nonumber \\ +
\frac{1}{2} \sqrt{F-2\alpha} \;, \quad
\end{eqnarray}
Clearly, for a positive $\alpha$, the central shell doesn't get
trapped, and the untrapped region around the center increases with
increasing $\alpha$, for both homogeneous and inhomogeneous models,
and are respectively illustrated in Fig.~(\ref{figure1}).

\section{Strength of singularity}
Finally, we need to determine the curvature strength of the naked
singularity, which is an important aspect of a singularity
\cite{ftp}.  A singularity is gravitationally strong or simply
strong if volume elements defined through Jacobi fields get crushed
to zero volume at the singularity, and weak otherwise
\cite{tce,PSJ,he}. It is widely believed that a space-time does not
admit an extension through a singularity if it is a strong curvature
singularity in the sense of Tipler \cite{tce, ft}. Clarke and
Kr\'{o}lak \cite{CCAK} have shown that in four dimensions a
sufficient condition for a strong curvature singularity as defined
by Tipler \cite{tce} is that for at least one nonspacelike geodesic
with affine parameter $\tau$, in the limiting approach to the
singularity, we must have
\begin{equation}
\lim_{\tau\rightarrow \tau_0}(\tau- \tau_0)^2 \psi =
\lim_{\tau\rightarrow \tau_0}(\tau- \tau_0)^2 R_{ab} K^{a}K^{b}
> 0 \label{eq:sc}
\end{equation}
where $R_{ab}$ is the Ricci tensor.  This provides a sufficient
condition for all the two-forms, defined along the singular null
geodesic, to vanish as the singularity is approached, and implies a
very powerful curvature growth establishing a strong curvature
singularity.

Following  \cite{djd}, we consider a timelike causal curve$ K^a ={d
x^a}/{d \tau}$ where $\tau$ is the proper time along particle
trajectory and  $K^a$ satisfies condition $K^a K_a = -1$. The radial
timelike geodesics must satisfy \cite{djd}:
\begin{equation}\label{geo2}
\frac{dK^t}{d\tau} + \frac{{\dot R}'}{R'} [(K^{t})^2-1] = 0.
\end{equation}
It has a simple solution $K^a = {d x^a}/{d\tau} = \delta^a_t, r=0$,
which is the worldline of the center of the collapsing cloud. In
terms of proper time we can describe it as
\begin{equation}\label{affine}
t_c(0) - t = \tau_0 - \tau.
\end{equation}
We consider the expansion of  $R$ near the cenet
\begin{equation}\label{expans}
R(t,r)=R_0(t) r + R_1(t) r^2 + \cdots
\end{equation}
where $R_0(t)$ is unity at $t=0$ and vanishes at $t=t_c(0)$. The
expression for $\psi$, with the help of eqs. (\ref{geo2}),
(\ref{affine}) and (\ref{expans}), becomes:
\begin{equation}\label{psi1}
\psi = -\frac{1}{\alpha} +  \frac{20 F_0}{[R_0(t)^4+8\alpha
F_0]^{3/2}}R_0^2 + { \cal O} {(R_0^2)}
\end{equation}
where ${ \cal O} {(R_0^2)}$ signifies terms which vanish faster than
$R_0(t)^2$ in the limit $R_0(t)\rightarrow 0$. Thus one finds that
$\lim_{\tau\rightarrow \tau_0}(\tau- \tau_0)^2 \psi =0 $ and
therefore the strong curvature condition is not satisfied. Thus the
Gauss-Bonnet term weakens the strength of singularity.

\section{discussion}
The low-energy expansion of supersymmetric string theory suggests
that the leading correction to Einstein action is given by
Gauss-Bonnet invariant. In this paper, in 5D, we have found exact
spherically symmetric LTB solutions, for the marginally bound case,
to Gauss-Bonnet extended Einstein equations (5D-LTB-EGB). This
describes gravitational collapse of spherically symmetric
inhomogeneous dust in a 5D space-time in EGB gravity.

The solution in turn is utilized to bring down the effect of the
Gauss-Bonnet term on the final fate of the 5D relativistic
gravitational collapse of a dust cloud. It may be noted that the
analogous 5D-LTB case exhibits critical behavior governing the
formation of black holes or naked singularities. The natural
questions would be, for instance, whether such solutions remain
naked with the correction terms of second order in the curvature? Do
they get covered? Does the nature of the singularity change in a
more fundamental theory preserving censorship?

We found that, as in the case of general relativity, a naked
singularity is inevitably formed. In the general relativistic case,
a naked singularity will form only when $M_0$ takes a sufficiently
small value, and therefore turning on the Gauss-Bonnet term worsens
the situation from the viewpoint of CCC. Our analysis shows that the
Gauss-Bonnet contribution has a profound influence on the nature of
the singularity and the whole picture of gravitational collapse
changes drastically.  While there may be an apparent horizon about
this singularity, for $\alpha> 0$, the singularity always remains
visible to any observer as the apparent horizon lies beyond
singularity which is actually not in the space-time. It is
interesting that the coupling constant of the Gauss-Bonnet terms
produces a change in the location of the apparent horizon by the
factor $2 \alpha$ is exactly the same as in the case of 5D null
fluid collapse in EGB.

The most interesting consequence of the second order curvature
corrections is that the final fate of gravitational collapse is
quite different in the sense that a massive naked singularity is
formed, which is disallowed in 5D-LTB. Thus we have shown here that
there exist regular initial data which lead to a massive naked
singularity violating CCC. However, since the strength singularity
is weaker as compared to the corresponding 5D-LTB, this may not be a
serious threat to CCC. According to Seifert conjecture \cite{sf} any
singularity that occurs, if a finite nonzero amount of matter tends
to collapse, into one point is always hidden. Hence, this is a
counterexample to Seifert conjecture as well. The singularities are
always naked as they are formed prior to the formation of apparent
horizon and there is no black hole formation at least for the
marginally bound case, and hence they must violate HC. Thus we have
a unique counterexample to all three conjectures. It would be
interesting to investigate further gravitational collapse in EGB
theory to see if these features are generic \cite{sj1}.  However, it
would be difficult to say that this is serious threat to CCC, since
the strength of singularity is weaker in 5D-LTB-EGB than in 5D-LTB.

It is seen here that the Gauss-Bonnet term modifies the time of
formation of singularities, and the time lag between singularity
formation and apparent horizon formation, in contrast to the 5D dust
models. Indeed, the time for the occurrence of the central shell
focusing singularity for the collapse is increased as compared to
the autologous 5D-LTB case. The reason may be, there is relatively
less mass-energy [see Eq.~(\ref{eq:m3})] collapsing in the
5D-LTB-EGB space-time as compared to the 5D-LTB case. In particular,
our results in the limit $\alpha \rightarrow 0$  reduce
\emph{vis-$\grave{a}$-vis} to 5D relativistic case.

\end{document}